\documentstyle[12pt]{article}
\begin{document}
\begin{center}
{\bf Cluster Hybrid Monte Carlo Simulation Algorithms}\\
\vspace{ 5mm}
{J. A. Plascak$^{1,2}$, Alan M. Ferrenberg$^{2,3}$ and D.P. Landau$^{2}$\\
$^{1}$Universidade Federal de Minas Gerais,
Departamento de Fisica - ICEx\\
C. P. 702, 30123-970, Belo Horizonte - MG  Brazil\\
$^{2}$Center for Simulational Physics,
The University of Georgia, Athens Georgia 30602\\
$^{3}$University Computing and Networking Services,
The University of Georgia, Athens Georgia 30602
}

\end{center}
\date{\today}

\begin{abstract}
We show that addition of Metropolis single spin-flips to the Wolff
cluster flipping Monte Carlo procedure leads to a dramatic {\bf increase}
in performance for the spin-1/2 Ising model. We also show that adding
Wolff cluster flipping to the Metropolis or heat bath algorithms in systems
where just cluster flipping  is not immediately obvious 
(such as the spin-3/2 Ising model)
can substantially {\bf reduce} the statistical errors of the simulations. 
A further advantage of
these  methods is that systematic errors introduced by the use of
imperfect random number generation may be largely healed  by hybridizing 
single spin-flips with cluster flipping. 
\end{abstract}
pacs{02.70.Uu,05.10.Ln,05.50.+q,05.70.Jk}

\section{Introduction}
The potential resolution of Monte Carlo (MC) computer simulations has
increased substantially over the past few years.\cite{landau0,Binder}
This has been due, in part, to the dramatic rise in the performance of
computers, but, more importantly, to the development of more powerful
data analysis and computer simulation techniques.\cite{Swendsen}
Histogram methods allow us to extract much more information from
simulation data than was previously possible.\cite{Swendsen,Histo} By
providing the ability to continuously vary the temperature or other intensive
parameters of a simulation, these techniques have greatly simplified the
analysis of simulation data by traditional means and, in addition, have
also played an important role in the development of new methods of
analyzing simulation data.\cite{highres,holm,kun} These methods are most
effective when very large numbers of spin configurations have been
generated, and it is the common belief that the number needed is
enlarged by correlations between successive states.\cite{errors,histerr}
More recently, a new generation of algorithms to calculate the density
of states accurately via a random walk in energy space have been devised
for producing canonical averages of thermodynamical quantities at essentially 
any temperature. \cite{murilo,fugao}
Simulation techniques have also improved immensely.  Fast
implementations of local update (Metropolis\cite{metrop}) algorithms have
been developed for a variety of models, while cluster-flipping algorithms
\cite{Swendsen,SW,Wolff}, which can dramatically reduce the correlation time
in a simulation, now exist for several classes of models.

A different approach to increasing the performance of computer
simulations is to combine several different algorithms into a single,
hybrid algorithm.  This idea is not new; hybrid Monte Carlo\cite{Duane},
Hybrid Molecular Dynamics\cite{Heermann}, Metropolis with
overrelaxation\cite{overrelax,creutz} and Multi-Hit
Swendsen-Wang\cite{MHSW} are some examples of hybrid algorithms.  In
these cases, however, the two algorithms that are combined perform the
simulation in different ensembles, either canonical/microcanonical or
canonical/fixed-cluster-distribution.  The approach we consider here is
to combine algorithms that work in the same ensemble, for our examples
the canonical ensemble, so that each of the individual component
algorithms is a self-sufficient simulation technique.  This eliminates
any concerns about how the mixing of ensembles could potentially
affect the quality or correctness of the results.  We will, however,
discuss the generalization of these ``proper'' hybrid algorithms to
include mixed-ensemble cases.

Our aim in  this work is two-fold. We first discuss, in the next section, 
the efficiency of a general hybrid algorithm and show how 
it can be improved in the case when 
Wolff plus  Metropolis is applied to the spin-1/2 two-dimensional Ising
model. Second, in section III, we apply a hybrid algorithm to the spin-3/2
two-dimensional Ising model for which a correct single cluster algorithm 
is not 
immediately obvious since the simple version does not take into account 
transitions
between states having different spin moduli (for instance, transitions
between $\pm 3/2$ and $\pm 1/2$ spin values). Further discussion and 
some concluding remarks are given in the last section.

\section{General Hybrid Algorithm: spin-1/2 Ising Model}

Consider a MC study of some model in which $N$ measurements of some
observable quantity $A$ (energy, magnetization, susceptibility,
cumulants, $etc.$) are made, and
for which there exist several different
algorithms that could be used to perform the simulation.  In order to
compare the efficiency of the different techniques, one needs to know
both the speed with which measurements are made and the degree to
which successive measurements are correlated.\cite{errors}  For this section,
we will define the efficiency $e$ for an algorithm as
$$
e = {{\rm \#\ of\ measurements\ generated\ per\ second} \over {2\tau_{A} +
 1}}~,
$$
where the integrated autocorrelation time $\tau_{A}$ is given by
$$
\tau_{A} = \sum\limits_{t=1}^{t=N} \left ( 1 - {t \over N} \right )
 \phi_{A}(t)~,
$$
with the time-displaced correlation function $\phi_{A}(t)$ for the
quantity $A$ calculated as
$$
\phi_{A}(t) = { {<A(0)A(t)> - <A>^2} \over {<A^2> - <A>^2} }~.
$$
Note that the correlation time, and therefore the efficiency of an
algorithm, can depend strongly on the particular quantity $A$ measured.

Now consider a hybrid simulation algorithm that combines several different
component algorithms.  To set up some notation, let $a$ represent the number
of different algorithms used, $N_i$ the number of measurements made with
simulation technique $i$ and $t_i$ the time (in seconds) required for
performing the update and making a measurement for technique $i$.
This time will, of course, depend strongly on the implementation of the
algorithms and the particular computers on which they are run.

The time in seconds needed to produce a measurement using the
hybrid algorithm is
$$
{ {\sum\limits_{i = 1}^{a} N_i t_i} \over
      {\sum\limits_{i = 1}^{a} N_i} }~,
$$
so that the efficiency of the hybrid algorithm becomes
$$
e = {\sum\limits_{i = 1}^{a} N_i \over \left ( 2\tau_{A} + 1 \right )
 \times
\left (\sum\limits_{i = 1}^{a} N_i t_i \right ) }~.
$$
Please note that the correlation time, and therefore the efficiency,
of the algorithm will depend on its specific implementation.  For
example, in a hybrid algorithm consisting of two components, $1$ and
$2$, the correlation time for the sequence $122122122122...$ would,
most likely, be different from the sequence $112222112222.....$ Which
of the two would produce the smaller correlation time would depend on
the dynamics (kinetics) of the individual algorithms.

We now demonstrate the development of the above hybrid algorithms 
by considering a
specific example, the spin -1/2 nearest-neighbor square-lattice 
Ising model at
its critical temperature, $T_c$. 
The Ising model has
 traditionally been used to test new simulation
algorithms and data analysis techniques because of its simplicity and
the exact finite and infinite-system solutions in the two-dimensional
model.\cite{Ising2d,onsager}  Because of the large amount of work done with
 the Ising model,
there exist several different simulation algorithms for it.  These can
be broken up into two major classes:  1) single-spin update
algorithms, including Metropolis, heat bath and microcanonical
algorithms, and 2) cluster algorithms, including the Swendsen-Wang\cite{SW}
 and
Wolff algorithms.\cite{Wolff}  We will concentrate on two of these
 algorithms\cite{Note}:
Metropolis\cite{metrop} and Wolff\cite{Wolff}.

Each of these algorithms has its strengths and weaknesses.  The
Metropolis algorithm is very efficient at equilibrating short-range
fluctuations in the system, and there exist highly-optimized multi-spin
coding implementations of the Metropolis algorithm.\cite{landau}
Unfortunately, the Metropolis method is not efficient at decorrelating
the long-range clusters that characterize the behavior of the system
near the critical point.  The Wolff algorithm, on the other hand,
concentrates its effort on the large clusters leading to greatly reduced
correlation times and a much smaller dynamic critical exponent $z$.
However, smaller-scale structures in the system, in particular regions
of disorder, are not handled efficiently by the Wolff algorithm. The
speed of the Wolff algorithm, based on the number of spins updated per
second, is also lower for Wolff than for multi-spin coding
implementations of Metropolis.  Because of the Wolff algorithm's smaller
dynamic exponent, it is clear that it will become more efficient than
Metropolis for sufficiently large lattices;  however, ``sufficiently
large'' might well be larger than the range of sizes of interest in a
particular study.  Work by Ito and Kohring\cite{ItoKohring} estimates
that Metropolis remains more efficient than Wolff, in terms of
independent measurements per second, for system sizes as large as $L =
70$ in two dimensions and $L = 100$ in three dimensions (running on a
scalar workstation).  This is, of course, strongly dependent on the type
of computer and the particular implementation of the algorithms used.
For example, with the programs, algorithms and computers used in this
study, we estimate that Wolff becomes more efficient than Metropolis for
$L \approx 32$ for $d = 2$ and $L \approx 16$ for $d = 3$.

Another concern with the Wolff algorithm is its sensitivity to flaws in
the random number generator used in the simulation.  Small but
significant systematic deviations from the exactly-known answer in the
$d=2$ Ising model and other systems have been reported and
investigated\cite{hoogland}-\cite{heuer}
using a variety of popular random number
generators.\cite{Marsag}-\cite{swc}  While the results of any
simulation method can be biased by subtle correlations in the random
numbers\cite{SantaBarbara}, the Wolff algorithm was found to be
particularly susceptible.  Despite these concerns about the Wolff algorithm,
 the dramatic reduction
in the correlation time is a very tantalizing effect.  If the speed and
efficiency at equilibrating small-scale structures of the Metropolis
algorithm is combined with the strength in decorrelating large-scale
structures of the Wolff algorithm, the resulting hybrid algorithm could,
in fact, be more efficient than either Metropolis or Wolff individually.

To test this possibility, we implemented a scalar hybrid algorithm which
combines the Metropolis and Wolff algorithms.  Simulations were
performed on IBM RISC/6000, DEC Alpha, PC Linux and SGI Power Challenge
workstations.  The spins were stored as bit variables, with up to 32
spin variables packed into a single computer word.\cite{landau}  Note that
 the
Metropolis algorithm can take advantage of this packing arrangement by
effectively updating many spins in parallel using multi-spin coding
techniques.  This will result in substantial improvement in performance
for increasing system size until all 32 bits are filled (for $L \ge 64$
in this implementation).  While the Wolff algorithm cannot make full use
of the multi-spin coding, it does benefit from the smaller memory
requirements of the packed-spin representation.  (Smaller memory means
that more of the system can be stored in the computer's cache memory
which results in much better performance.)

The random number generators used for the simulation must be chosen with
great care, especially for the Wolff algorithm.\cite{rngerror}  After
 performing
extensive tests of several generators, we selected the following as being
the fastest random number generators that would give us the correct
answer within the precision of our testing.  For the Wolff algorithm, we
 used a combination generator by
l'Ecuyer\cite{ecuyer} that was recommended as a ``perfect'' random number
 generator
in the Numerical Recipes column in Computers in Physics.\cite{numrep}  With
 this
program, we can produce a random number in 840.2 nsec on an SGI Power
Challenge workstation with a 194 MHz R10000 processor.  For the
Metropolis part of the simulation, we used a faster, shift-register
generator, R1279, which can produce a random number in 21.4
 nsec.\cite{rngcomment}

To see how the hybrid algorithm behaves when poor random numbers are
used, we ran a series of simulations deliberately using  a bad random
number generator for the Wolff algorithm.  We thus used the R250
shift-register generator\cite{shiftreg} which is known to introduce
significant systematic errors for the d=2 Ising model.\cite{rngerror}  
We performed
hybrid updates consisting of one Wolff update followed by 0, 1, 2, 3 and
4 Metropolis updates.  (A simulation consisting of only Metropolis was
also performed for completeness.)  For each hybrid, 16 independent
simulations consisting of $3 \times 10^6$ hybrid steps were performed.
The results for $L=16$ are shown in Fig. \ref{Fig1} for the energy and specific heat.
For the internal energy the Wolff algorithm yields the
wrong answer by an amount which is more than 35 times the calculated
error bar.  With the inclusion of $50\%$ of Metropolis flips this error
is reduced by a factor of 10, and with $80\%$ Metropolis flips, no
discernible error is seen.  Very much the same behavior is seen in the
specific heat, although the rate of convergence to the correct answer is
slightly different.

Not only are the results like to be more correct if the different
flipping mechanisms are mixed, but the performance is also improved.
For the magnetization the relative efficiency of the hybrid algorithm,
with $50-80\%$ Metropolis flips added, is about $30\%$ greater than for
Wolff alone, as can be seen from Fig. \ref{Fig2}.  It is surprising that even for $L=64$, where
the Metropolis algorithm is much less efficient than the Wolff
algorithm, the hybrid is significantly {\bf more} efficient.  Although
for pure Metropolis, the relative performance becomes markedly worse as
the lattice size increases, the same is not true for the hybrid
algorithm.  For the internal energy the relative efficiency, also shown in
Fig. \ref{Fig2}, is much better still for the hybrid algorithm, by more than a
factor of two.

\section{Hybrid algorithm: spin-3/2 Ising model}

For models  with higher values of spin, not only are Monte Carlo simulations, as 
well as specific algorithms, less ubiquitous than for their two-state counterpart,
but no exact solution is still available for their critical temperatures.
Thus, the basic ideas of the last section need to be extended to 
more general models, e.g. the spin-$3/2$ Ising model, where 
each spin state can assume values $\pm 3/2,~\pm1/2$. 
Although some spin-$1$  \cite{jain}-\cite{tsypin} and spin-$3/2$ 
\cite{cesar}-\cite{pla} models have already been studied through 
Metropolis technique and cluster spin-flipping \cite{carlos}, 
there is still a lack of
a  detailed analysis of the statistical and systematic errors even in 
the simple Ising limit. So, before starting to implement a hybrid
algorithm to this model it is interesting to see first what one gets with
single-spin flipping procedures.

To analyse the statistical errors of some observable thermodynamic
quantity $A$ we first applied just the Metropolis algorithm to the
spin-$3/2$ Ising model.  We ran $6.02\times 10^6$ Monte Carlo steps 
(MCS) per spin with
$2\times 10^4$  configurations discarded for thermalization 
on different lattice sizes $L$ ($8\le L \le 128$) and using the
``perfect'' random-number generator.\cite{ecuyer} We measured the
energy $E$, magnetization $M$, fourth-order cumulant of the magnetization 
$U$ and the quadrupole moment $Q$ (the mean value of the square of the spins). 
Typical results of the relative error $\Delta A/<A>$ 
for different lattice sizes $L$ are shown in Fig. \ref{Fig5}
at $t=k_BT/J=3.29$, a value close to the critical temperature.
Using 'coarse-graining' \cite{landau0}
we have estimated $\Delta A$ through
$$(\Delta A)^2=(<A^2>-<A>^2)/N~,$$
for large enough $N$,
where we divided our data into $N=MCS/n$ bins of different lengths $n$
($n$ ranging from $5$ to $10^5$). 
The relative error in the magnetization and its cumulant increases as $L$ increases
while for the energy and the quadrupole moment it stays almost constant.
In terms of different degrees of self-averaging, $M$ and $U$ are
non-self-averaging while $E$ and $Q$ exhibit a lack of self-averaging
(the number of `effectively' independent measurements through the computation 
of the correlation time $\tau$ is certainly
necessary \cite{errors} for a more detailed analysis of the errors. This is,
however, outside of the scope of the present work).
We have also noted no significant changes in the errors 
by using different random number generators, even taking the poorer congruential
one, and the data are also depicted Fig. \ref{Fig5}. 
Fig. \ref{Fig6} shows the results
of the magnetization cumulant $U$ as a function of temperature for
different MCS for the lattice size $L=128$. Here, we have used the histogram
technique at $t_o=3.29$ in order to obtain estimates for other values
of temperatures close to $t_o$. It is worthwhile to analyse 
such behavior since we will use the crossings of the cumulants $U$ to 
locate the critical 
temperature of the present model. Besides having large error bars one can 
see that the mean value of the cumulant is strongly dependent on the number
of MCS used to obtain the statistics. The dependence with the number
of states generated for this lattice size is more pronounced by using the 'perfect' 
random generator
(note that the mean values of $U$ with $6\times 10^6$ MCS and 'perfect' generator 
are comparable to those with $3\times 10^6$ MCS and the congruential one)
although both  converge to the same limit as the number of
MCS gets very large. Within the error bars we also notice almost no
systematic error due to the use of different random number generators,
in contrast to the case of 
the Wolff algorithm which, with a bad random number generator for the spin-1/2 
model, gives wrong results for the energy and specific heat
(see Fig. \ref{Fig1}). The same qualitative behavior
of Fig. \ref{Fig6} (large error bars and a strong dependence of the
cumulant with the number of MCS) is also seen for other lattice sizes $L$. Even by
substantially increasing the MCS per spin one still gets large errors,
mainly for the magnetization and its cumulant (see also Fig.
\ref{Fig5}).

We show in Fig. \ref{Fig7} the reduced pseudo-critical
temperature $t_c$ as a function of $L^{-1}$ (in fact $L^{-1/\nu}$, where $\nu =1$ 
for the two-dimensional Ising universality class) obtained from the crossings 
of the fourth-order cumulant of the magnetization for different values of $L$
using just the Metropolis algorithm. Each point in that figure represents
the crossing point of the cumulant $U_L$ of the lattice size $16\le L\le 128$ with the
corresponding cumulant of the smallest lattice $U_8$. Only a poor estimate of 
the critical reduced temperature can be achieved in this case which can be
ascribed to the large error bars obtained in computing $U_L$ as well as its
strong dependence on the number of MCS taken in the statistics. 
In particular, we have $t_c=3.288(1)$  with 'perfect' and $t_c=3.287(1)$  with 
congruential random number generators which are, even so, comparable to the 
more recent series expansion result $t_c=3.2878(22)$.\cite{jensen}
We can see that, in general, no systematic error due to
random number generator is observed for the Metropolis algorithm. Moreover, 
within the error bars, very similar results are also obtained by running the 
symmetric heat bath single spin-flip procedure.

It is clear that one way to improve the
accuracy of the location of the critical temperature with Metropolis can be done
by increasing the MCS in order to achieve better statistics.
This will require, of course,  much more computer time, mainly for large
lattice sizes. We can,
however, use the results of the previous section in order to
construct a hybrid algorithm where, with not much extra computer
time,  more precise results could be obtained. The first step is thus to
implement a Wolff algorithm for this model. In a straightforward way, this 
implementation can be done by activating bonds between parallel 
nearest-neighbors spins $S_i$ and $S_j$  according to the probability
$p(K_{ij})=1-\exp (-2K_{ij}S_iS_j)$ and, when the full
cluster has been activated, all its spins are reversed. This procedure
has, however, two main differences regarding the spin-1/2 systems which
we have to keep in mind:
i)  now, the probability $p(K_{ij})$ depends on the particular configuration of 
    the parallel spins and can have three possibilities, depending on
    whether $\{S_iS_j\}$ are $\{{{3} \over {2}} {{3} \over {2}}\}$, 
    $\{{{3} \over {2}} {{1} \over {2}}\}$ and 
    $\{{{1} \over {2}} {{1} \over {2}}\}$ (and also the corresponding reversed
    configurations);
ii) this procedure  alone is not ergodic in the sense that it does not take into account 
    transitions between spin states with different spin magnitudes (it keeps
    fixed the number of $\pm {{3} \over {2}}$ and $\pm {{1} \over {2}}$ spins in
    each configuration and, for instance, the quadrupole $Q=\sum _i S_i^2$ is 
    always a constant).
While the former is just a generalization of the bond probabilities 
activation for systems with more degrees of freedom, the latter is really a 
problem since we can not generate all possible configurations for the model.
A mixed cluster algorithm has already been proposed to overcome such a 
non-ergodicity in the case of the spin-1 Blume-Emery-Griffths model.\cite{carlos}
However, a natural hybridization procedure, based on the discussion of
the last section, and also from embedding algorithms \cite{brower} proposed 
to study of spin-1 models \cite{blote} can be worked out here by simply
alternating one Metropolis sweep with $p$ Wolff steps where 
$p$, in principle, can depend on the system size. The inclusion of alternate single spin-flip 
sweeps will make this hybrid algorithm ergodic and much simpler than a possible
generalization of the mixed cluster procedure to the present spin-$3/2$
model. In order to test the efficiency
of this hybrid algorithm we have done extensive simulations for the $L=4$
lattice where we can compare the results of the simulations with the exact 
ones. We ran a total of $1.2\times 10^7$ hybrid MCS per spins each one including 
$p$ Wolffs intercalated by one Metropolis sweep. The results are shown
in Fig. \ref{Fig8} for the 'perfect' random generator. One can readily see
that all the results are in general compatible to the exact ones 
within the error bars.
However, by including some Wolff steps the mean values initially oscillate
for small $p$, have  a better agreement for $p\sim 5$ and finally deviate
for large $p$. We also note that the errors are almost the same for
$p=0$ and $p=9$ and are slightly smaller around $p\sim 5$. The slight
deviation of the error as a function of $p$ reflects the fact that we have a 
reasonable number of MCS per spin to obtain a good statistics for this 
small lattice, even with just the Metropolis algorithm (this is not the case
for larger values of $L$, as we shall see below). Moreover, the worse 
results for large values of $p$
can be ascribed to the non-ergodicity of the present simple Wolff algorithm
(nothing is gained by increasing the number of Wolff steps since we get
stuck in the configurations having the same number of $\pm {{3} \over {2}}$ 
and $\pm {{1} \over {2}}$ spins).
The overall picture then suggests the use of this hybrid algorithm with
$p\sim 5$ (although $p$ can also vary with $L$). In order to test this 
assumption we applied this procedure to the $L=128$ lattice (and close to
the critical temperature) and obtained the magnetization cumulant $U$ with $5$
Wolff steps. The corresponding results are also shown in Fig. \ref{Fig6}.
There is, in this case, no sensitive difference in the data by taking
$3\times 10^6$ or $6\times 10^6$ MCS with the hybrid algorithm. Surprisingly,
the statistical errors are now almost two orders of magnitude smaller than
those with just the Metropolis algorithm (the errors in Fig. \ref{Fig6}
for $L=128$ are in fact much smaller than the symbol sizes). 
To get this same precision
with only Metropolis one would have to compute an order of $10^8$
configurations for this lattice size. The relative error of the energy, 
magnetization and its cumulant, and the quadrupole for other values of
$L$ are shown in Fig. \ref{Fig9}. While now they are almost constant for
$M$ and $U$ (exhibiting lack of self-averaging), they decrease for $E$ and $Q$
(behaving now as self-averaging quantities). It is also important to notice
that the hybridization process is again almost insensitive to the quality of 
the random number generator. The above results strongly indicate 
$p=5$ as a good trial also for other lattice sizes.

In Fig. \ref{Fig10} we present  the reduced pseudo-critical
temperature $t_c$ as a function of $L^{-1}$ obtained from the hybrid
algorithm described above with the 
'perfect' random generator through the crossings of the fourth-order
cumulant of the magnetization $U_L$. It was possible, in this case, to get a good
resolution from crossings of $24\le L \le 128$ with three
different smaller lattices, namely $U_8$, $U_{12}$ and $U_{16}$.
The quality of the results are now apparent and yields the extrapolated value
$t_c=3.28799(7)$. Just for completeness, in Fig. \ref{Fig10} we also give 
the corresponding values by taking the congruential generator with the
present hybrid algorithm. As it is faster, we were able to use, with the
same computing time, lattices as large as $L=192$ to get 
$t_c=3.28789(7)$. We have then, so far, the best estimate for the critical 
temperature of the two-dimensional spin-3/2 Ising model: $t_c=3.28794(7)$.

It is worthwhile now to address some comments regarding the universality
of these  models. Regardless the number of states  each spin can assume,
all $d$-dimensional systems are in the same (Ising) universality class. This fact is
apparent in Figs. \ref{Fig7} and \ref{Fig10},
(mainly the latter one) where the temperatures are all along a straight line
as a function of $L^{-1/\nu}$ with $\nu =1$  in two dimensions. However, two more
universal quantities can be readily observed from the present simulations. 
First, the magnetization fourth-order cumulant $U^*$
at the transition temperature can also be estimated from our data 
to give $U^*\simeq 0.612(1) $, a value expected for $d=2$ 
Ising systems undergoing a second-order phase transition. 
This result comes from Fig. \ref{Fig12} where each point was obtained
by fixing $t$ at our estimate $t_c$ and looking the cumulant there for
different lattices. Second, a  quantity
which is studied less often, is the
probability distribution of the magnetization $M_L$, 
$P^*(b_oL^{\beta /\nu}M_L)$ which, for large enough systems at the critical
temperature is a universal function.\cite{bruce1,binder2,bruce2} In this equation
$b_o$ is a non-universal constant chosen to give
a unity variance for the distribution $P^*$. 
Fig. \ref{Fig13} shows the fixed-point order parameter distribution for
the two-dimensional Ising universality class obtained from models
with spin-1/2,1,3/2 at the critical temperature and lattice size $L=32$. 
For each model  $10^7$ steps were performed with Metropolis algorithm and
using the R1279 random number generator.
The quality of this match clearly reveals the hallmark of the $P^*$
distribution for the Ising universality class.

\section{Discussion}
The results shown in the previous sections supply strong support for
the use of hybrid algorithms as a means of effectively speeding up
simulations and also improving the quality of the results.  
Another advantage is that efficient,
parallel implementation of the hybrid algorithm on distributed
memory machines is straightforward.  A Wolff process running on one
processor can ``feed" states to other processors which then perform
multiple Metropolis updates.  The number of Metropolis updates can be
varied to maximize load balancing.  Data are gathered together from
all states which have been generated and then used to construct
histograms.  This procedure can be enhanced still further by the inclusion
of microcanonical updates which require no random numbers!  One hybrid
update would then consist of, {\it e.g.} 1 Wolff update plus 5 Metropolis
updates plus 10 Microcanonical updates.

Although we have described hybrid algorithms for one of the simplest
models in statistical mechanics (Ising), we believe that the lessons drawn from
these studies will be more broadly applicable.  For example, continuous
spin systems may be (randomly) projected onto Ising models which can
be easily simulated using these hybrid algorithms.  Histogram analysis
of the data can also be used in a similar fashion to produce extremely high
resolution results.  Of course, the relative performance of each component
of the hybrid algorithm will depend upon the specific model, so that
``tuning" will be required for each study.  Furthermore, these single
ensemble hybrid methods can be combined with other algorithms to further
improve performance.  For the Ising model the microcanonical method is
extremely fast and can be easily included.  For a classical Heisenberg
model, the over-relaxation method provides an effective microcanonical
simulation component to a hybrid algorithm.

In summary, we have demonstrated that hybrid Monte Carlo spin-flip
 algorithms,
which include ``slower'' Metropolis steps,
can be made to be effectively faster than cluster flipping algorithms.
Furthermore, and perhaps more significantly, they yield substantially
more accurate results than does the simple Wolff algorithm (for the spin-1/2
model) or single Metropolis algorithm because the
alternation of updating methods breaks up random number correlations.

acknowledgments{

This research was supported in part by Conselho Nacional de Desenvolvimento
Cient\'\i fico e Tecnol\'ogico (CNPq-Brazilian Agency) and NSF grant 
DMR-0094422.
}

\newpage

\begin{figure}
\caption{Dependence of the estimate of the internal energy $E$ per particle in units
of the exchange interaction $J$ (circles) and the specific heat $c$ in units of 
$k_B/J^2$ (squares) with the
fraction of Metropolis spin-flips.  Results shown are for the d=2 spin-1/2 Ising
model on a $L \times L$ square lattice with $L=16$ at $T=T_c$.  The dashed lines
represent the exact solution. }
\label{Fig1}
\end{figure}

\begin{figure}
\caption{Variation of the relative efficiency (compared to pure Wolff) of
the Hybrid algorithm as measured from the results for the magnetization
$M$ (filled symbols) and for the energy $E$ (open symbols)
with the fraction of Metropolis spin-flips for different lattice sizes.
Results shown are for the d=2 spin-1/2 Ising model at $T=T_c$ on $L \times L$
lattices. Where not shown, the error bars are smaller than the symbol size.}
\label{Fig2}
\end{figure}

\begin{figure}
\caption{
Relative error of the magnetization $M$, its cumulant $U$, energy $E$
and quadrupole $Q$ as a function of lattice size $L$ for the $d=2$ spin-3/2
Ising model. Filled symbols were
taken with the ``perfect'' random-number generator \cite{ecuyer} while
the empty ones with the congruential generator. Full lines and dashed
lines are guide to the eyes.
}
\label{Fig5}
\end{figure}

\begin{figure}
\caption{ 
Magnetization cumulant $U$ as a function of reduced temperature $t$
for the  $d=2$ spin-3/2 Ising model with $L=128$. 
All data were obtained from histograms taken at $t=3.29$
(close to the critical temperature). The numbers in the legends stand 
for MCS. Open squares and open diamonds
are the results for Metropolis with congruential random number generator 
(ME-C). Full squares and full diamonds with the 'perfect' generator (ME-P).
Open circles and full circles are the results for the hybrid algorithm
with 5 Wolff steps using the congruential (H5-C) and 'perfect' (H5-P) generators, 
respectively (these data are almost collapsed within the resolution of
this Figure). The magnitude of the error bars with the Metropolis algorithm 
for $3\times 10^6$ and $6\times 10^6$ MCS are indicated. For the hybrid 
algorithm the errors are much smaller than the corresponding symbol sizes.
}
\label{Fig6}
\end{figure}

\begin{figure}
\caption{Pseudo-critical temperature $t_c$ as a function of the inverse of
lattice size $L^{-1}$. Results obtained from the crossings of the fourth-order
cumulant of the magnetization using just the Metropolis algorithm with 
different $L$ for the  $d=2$ spin-3/2 Ising model. 
Circles are the results with 'perfect'
random generator (ME-P) and squares with congruential (ME-C). 
For clarity, the errors in the
congruential data are not shown (they are, however, of the same order as in the 
'perfect' case).  
}
\label{Fig7}
\end{figure}

\begin{figure}
\caption{ 
Magnetization $M/M_o$, energy $E/E_o$, quadrupole $Q/Q_o$ and magnetization cumulant
$U$ as a function of Wolff $p$ steps for the $L=4$ lattice at $t=3.0$. 
$M_o$, $E_o$, and $Q_o$ are the corresponding saturated values at $t=0$.
Results for the  $d=2$ spin-3/2 Ising model. 
The dashed line represents the exact solution.
}
\label{Fig8}
\end{figure}

\begin{figure}
\caption{ Relative error as a function of lattice size for the magnetization
$M$ (squares), its cumulant $U$ (circles), energy $E$ (diamonds) and quadrupole 
$Q$ (triangles) with the hybrid algorithm and $p=5$. Filled symbols have been obtained by
using the congruential generator and open symbols by using the 'perfect' one in 
the  $d=2$ spin-3/2 Ising model. The lines are guide for the eyes.
}
\label{Fig9}
\end{figure}

\begin{figure}
\caption{Pseudo-critical temperature $t_c$ as a function of the inverse of
lattice size obtained from the crossings of the fourth-order cumulant
of the magnetization with different $L$. The three sets of the filled symbols
have been obtained according to the hybrid
algorithm with `perfect' random number generator by considering the crossings of
$U_{24\le L \le 128}$ with $U_8$, $U_{12}$ and $U_{16}$, respectively.
The three sets of the open
symbols have been obtained according to the hybrid
algorithm with congruential random number generator by considering the crossings of
$U_{24\le L \le 192}$ with $U_8$, $U_{12}$ and $U_{16}$, respectively.
}
\label{Fig10}
\end{figure}

\begin{figure}
\caption{Estimate of the fourth-order cumulant value at the transition
$U^*$. The straight line  corresponds to a linear fit of the data.
}
\label{Fig12}
\end{figure}

\begin{figure}
\caption{$P^*$ as a function of $b_oL^{\beta /\nu}M_L$ at $T_c$ for the
spin-1/2,1,3/2 Ising models with lattice size $L=32$. The simulations
have been done at the exact value of $T_c$ for spin-1/2, $T_c=1.6935$
obtained from series expansions for spin-1 \cite{adler} and our present result
for spin-3/2.
}
\label{Fig13}
\end{figure}


\begin{thebibliography}{99}
\bibitem{Note} Note that for the Swendsen-Wang and Metropolis
algorithms, one update means one complete update of the lattice (MCS);
in the Wolff algorithm, one update is less than one MCS and depends on
the temperature.  For simulations at $T_c$, and the system sizes
considered in this study, a Wolff update is approximately 0.55~MCS.
\bibitem{rngcomment} Because of the sensitivity of the Wolff algorithm
to small imperfections in the random number generator, it is vital to
use a very good generator.  In addition, the L'Ecuyer generator produces one
 number
per call, while our implementation of R1279 produces an entire vector of
numbers.  This reduces the computing overhead and thus makes the R1279
generator even faster in comparison.
\end{thebibliography}
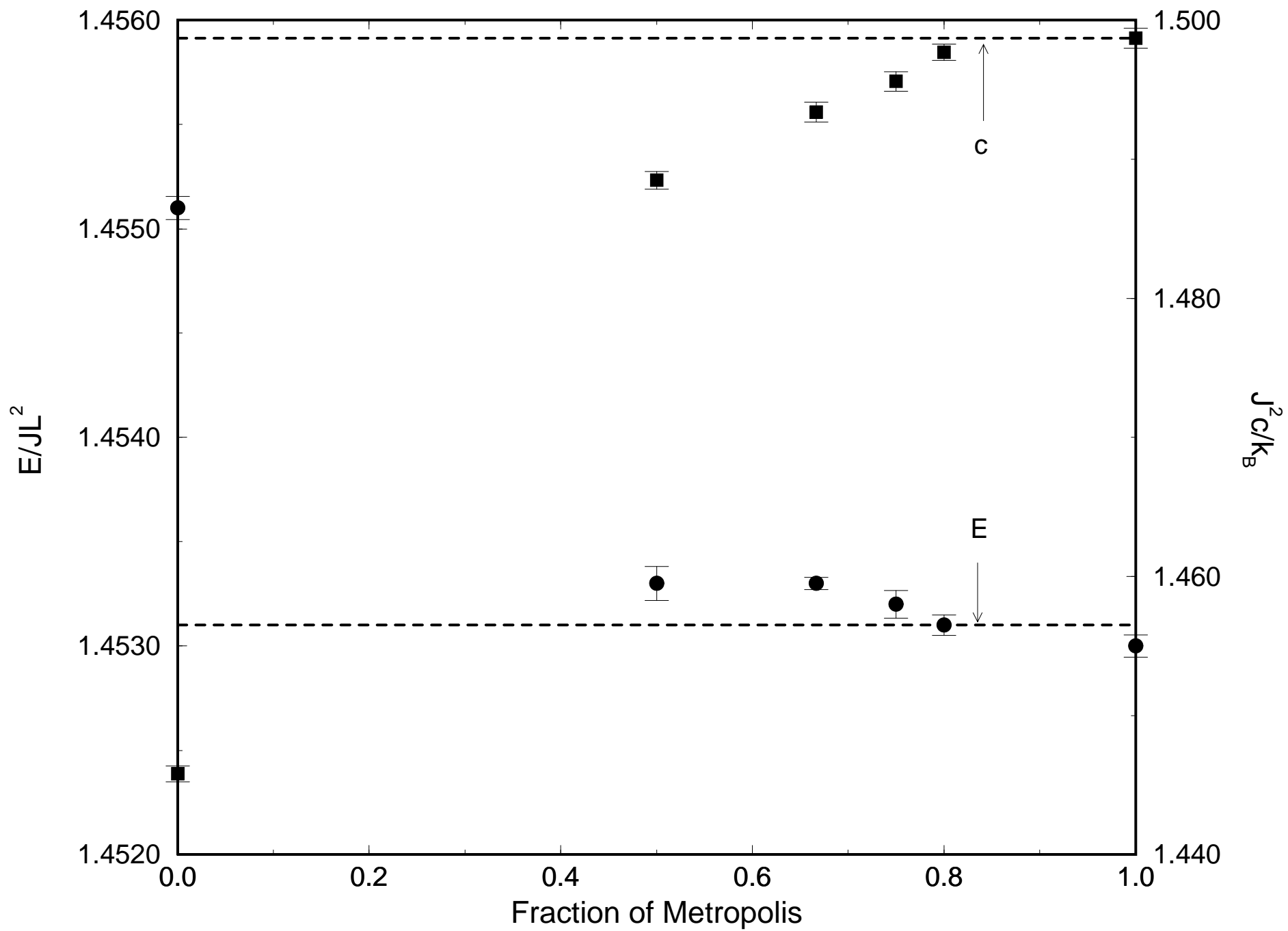
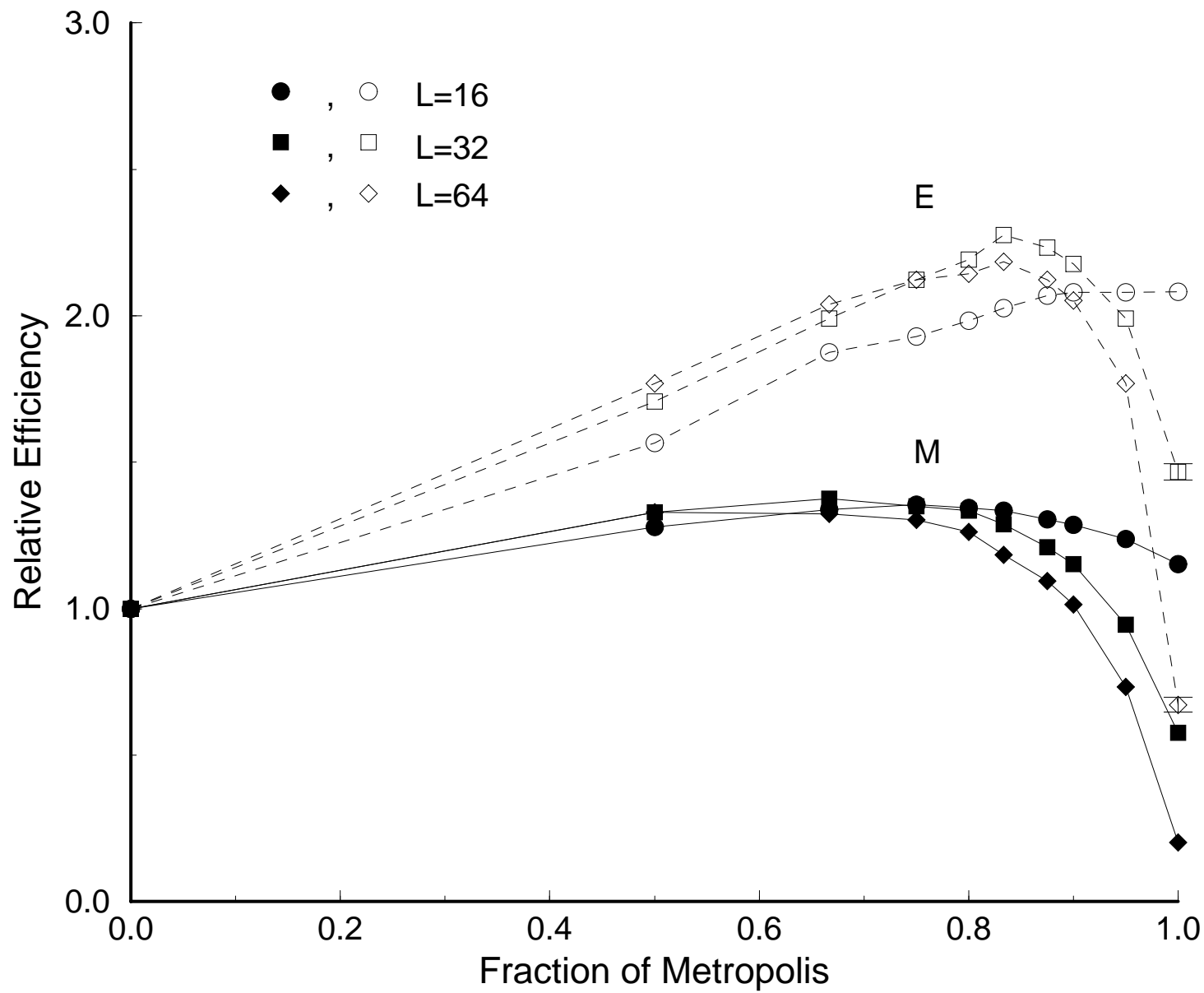
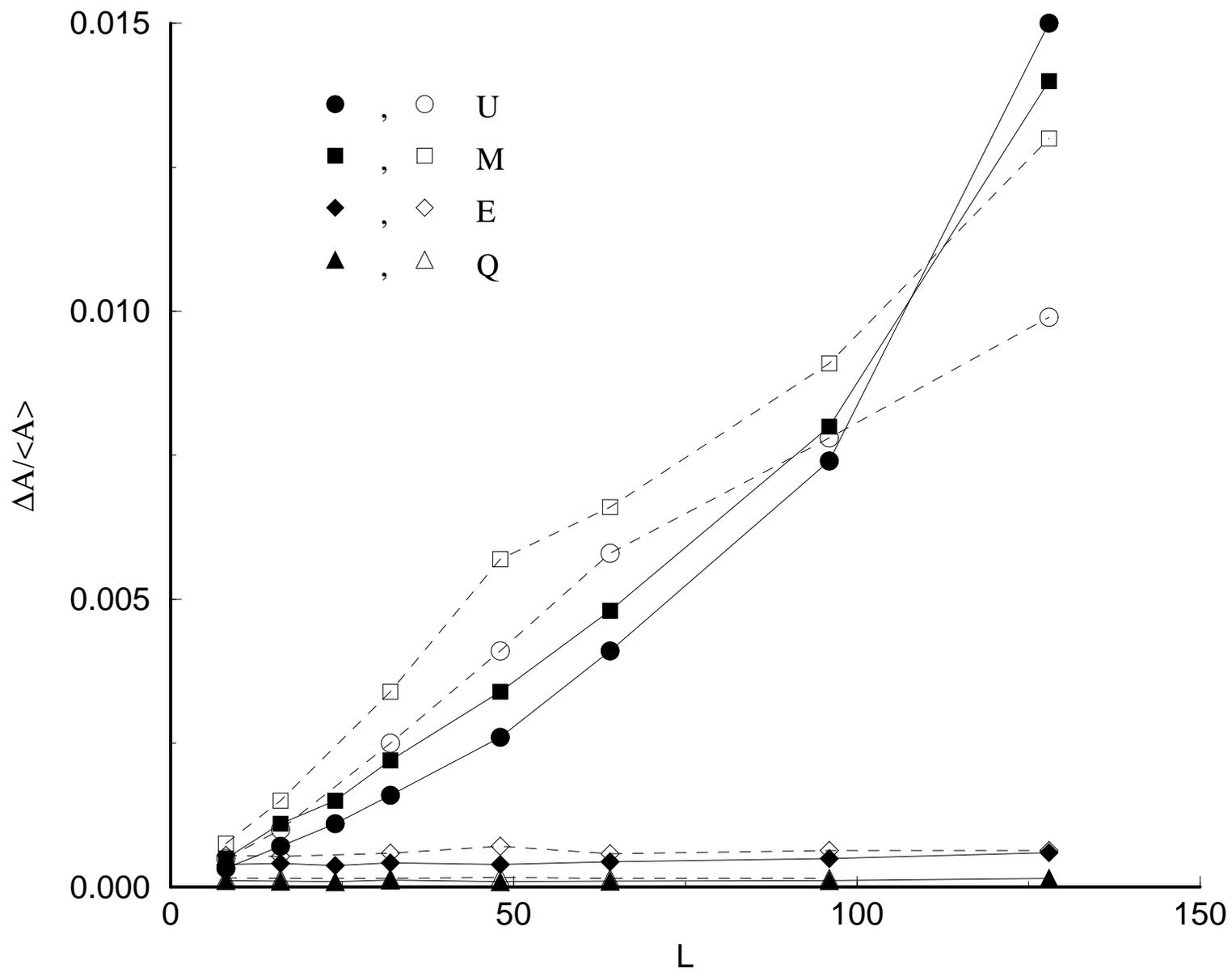
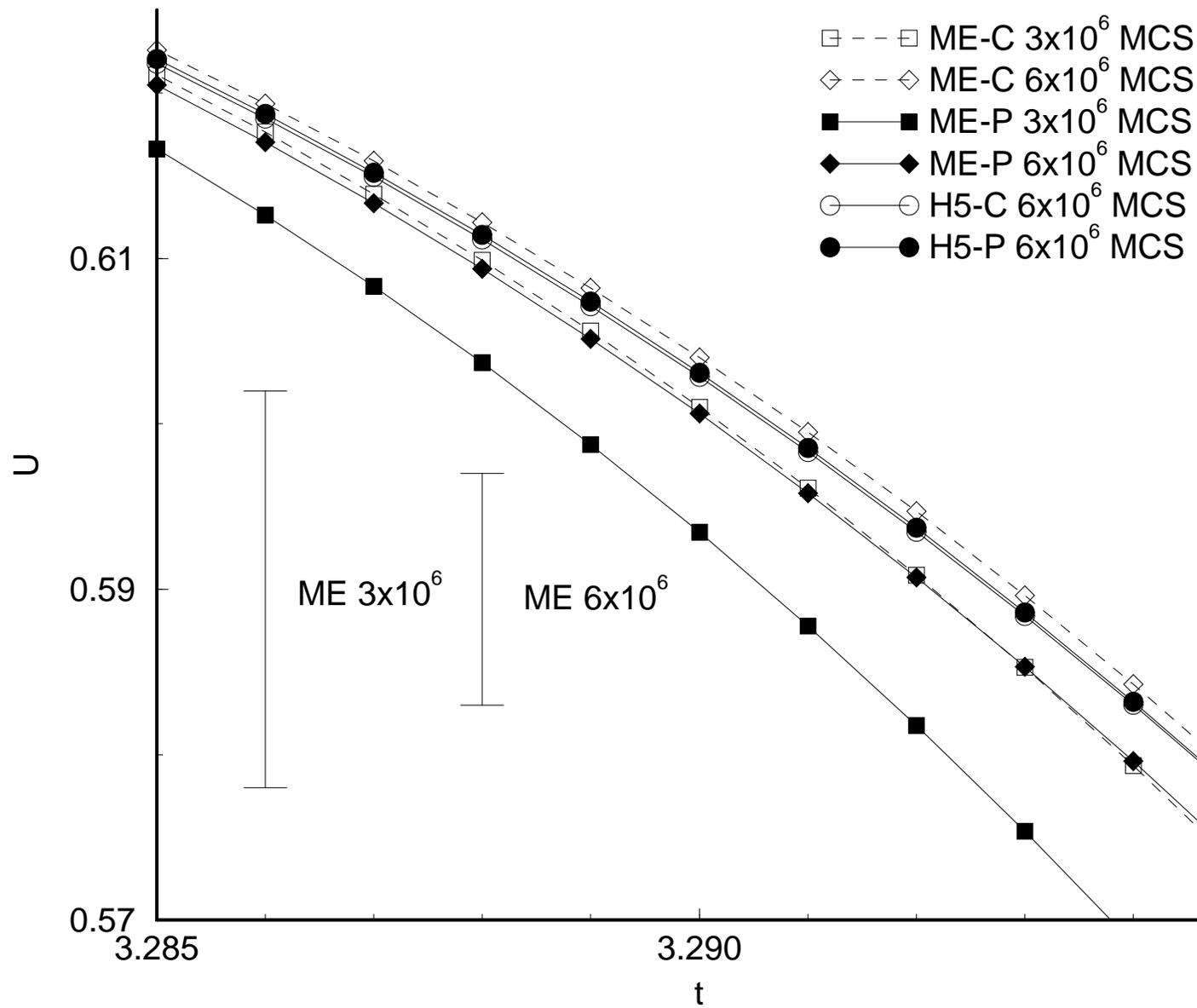
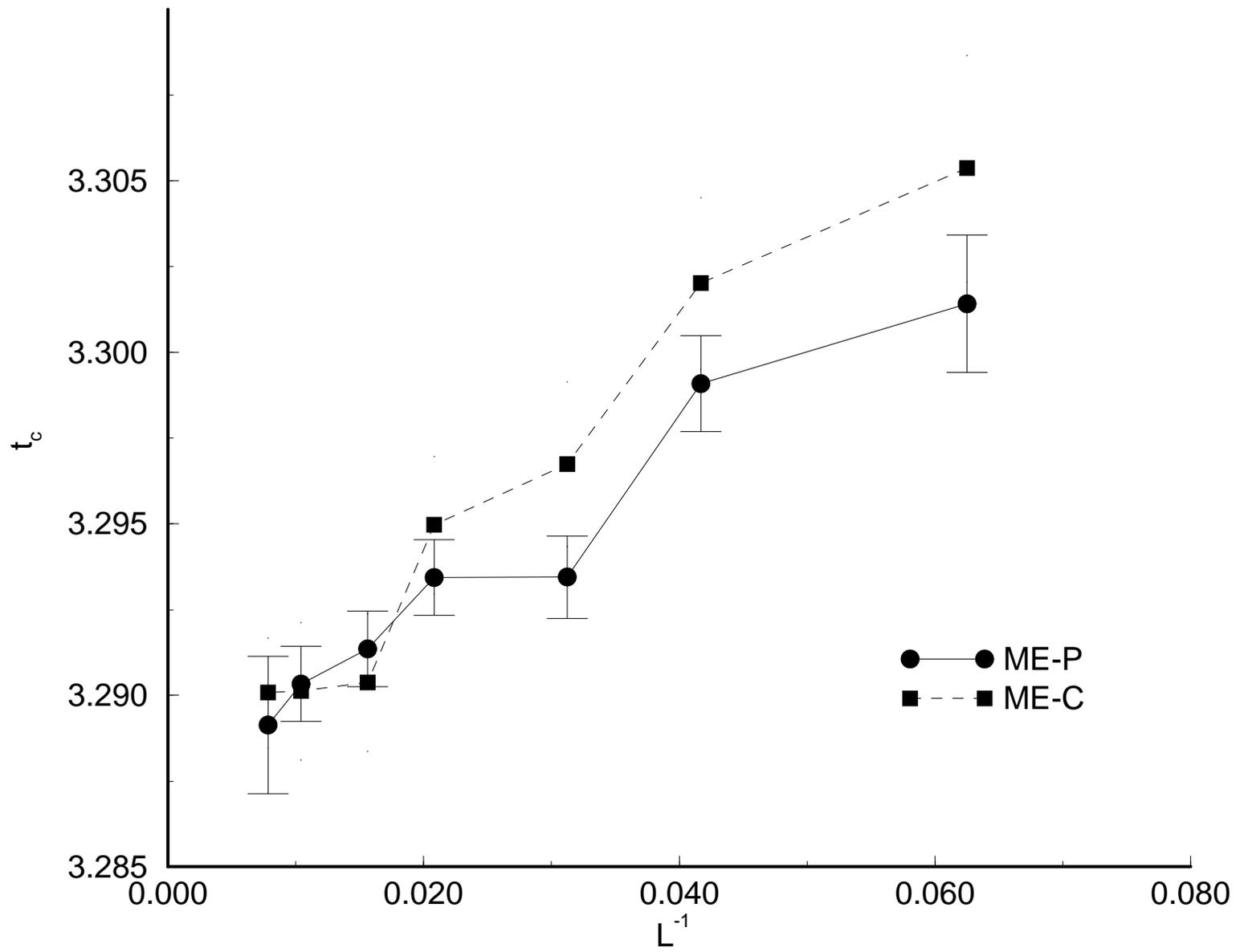
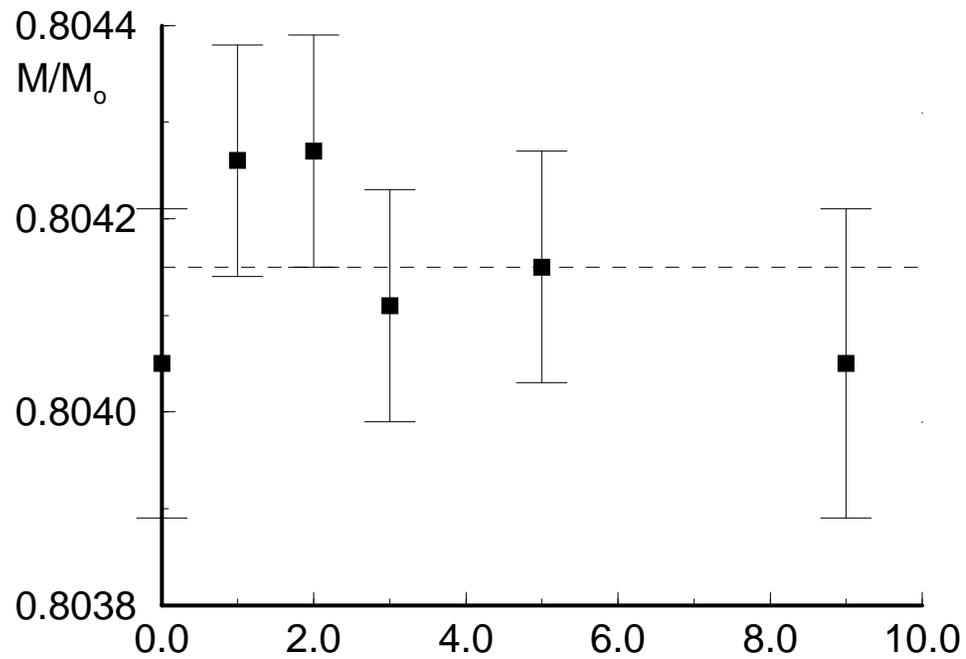
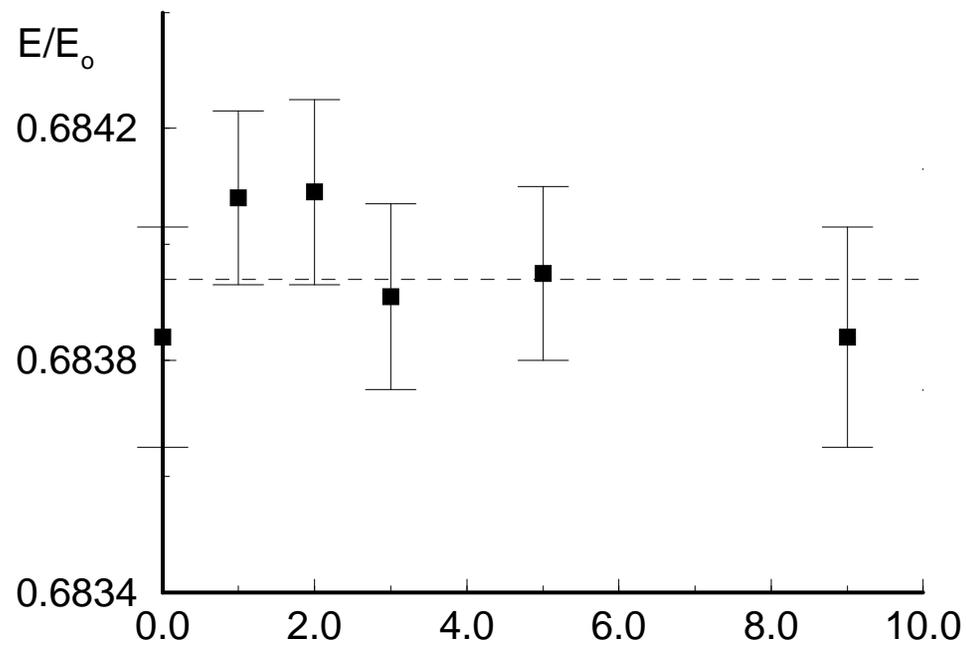
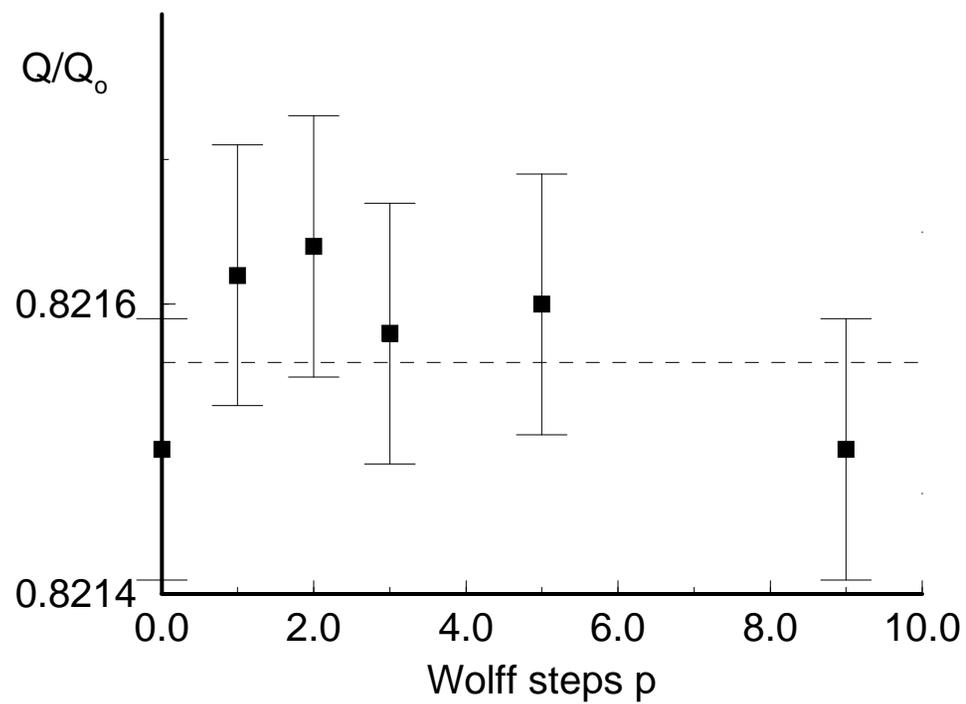
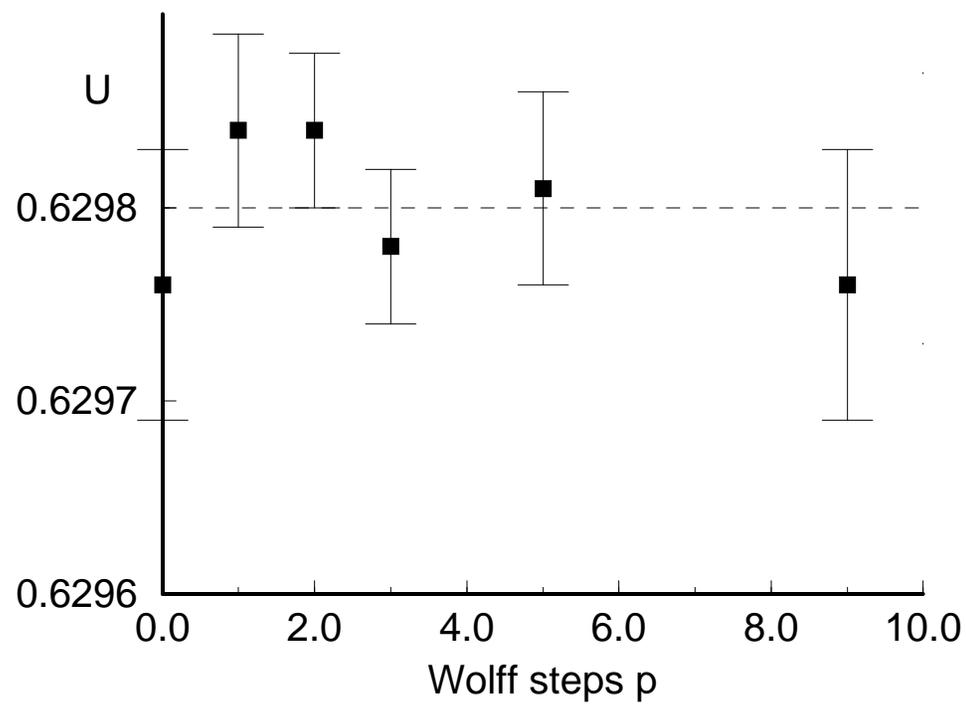
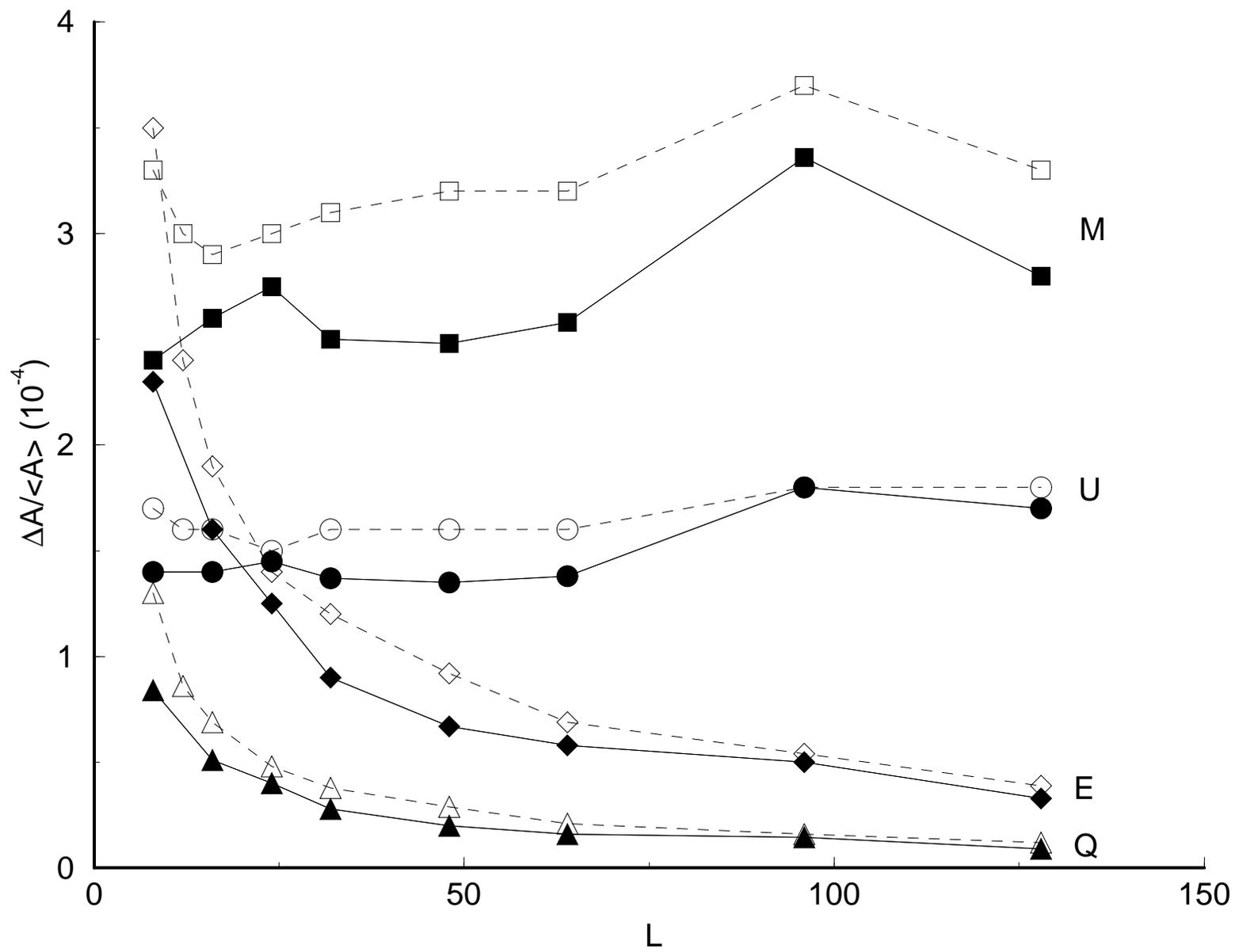
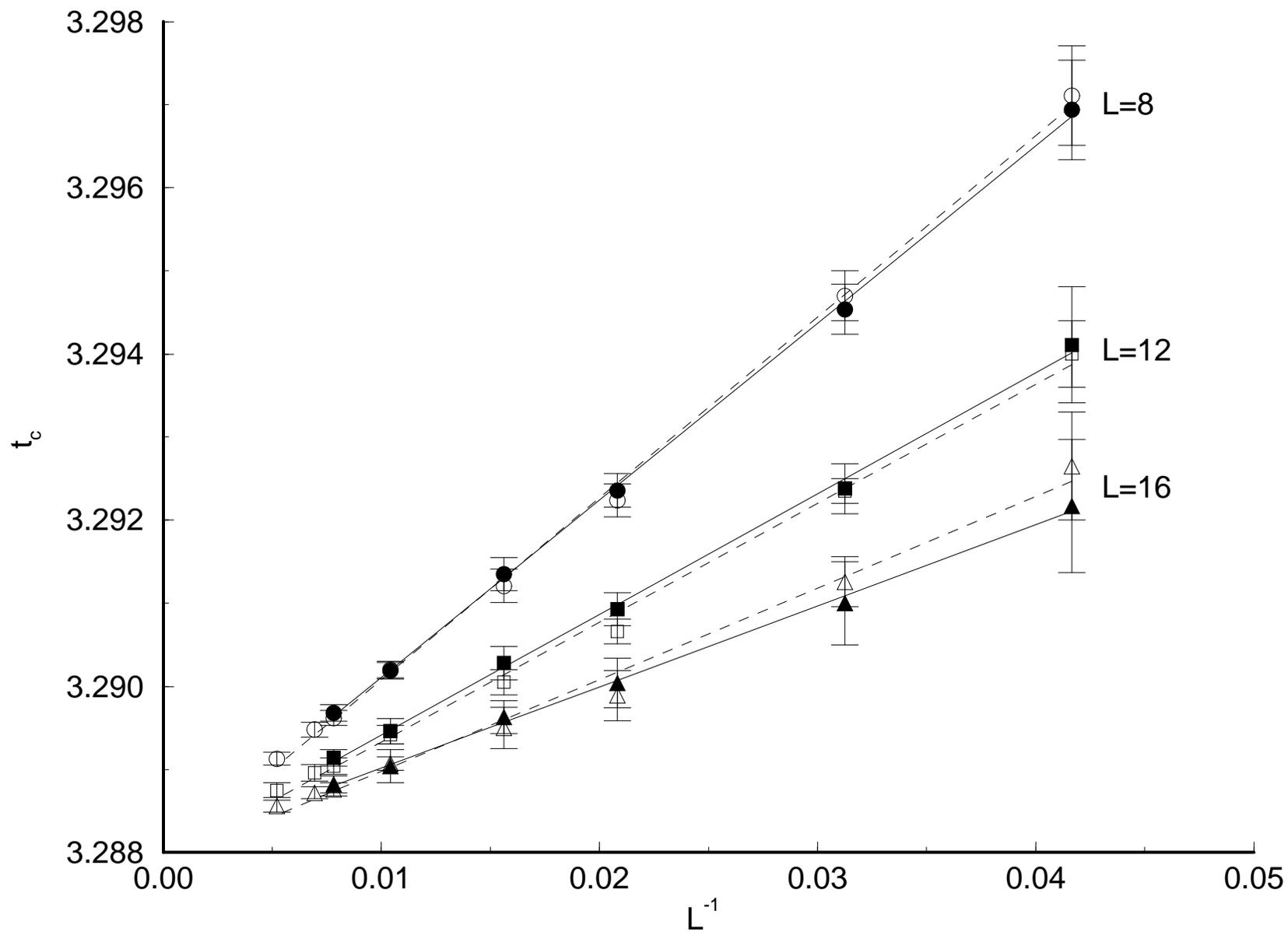
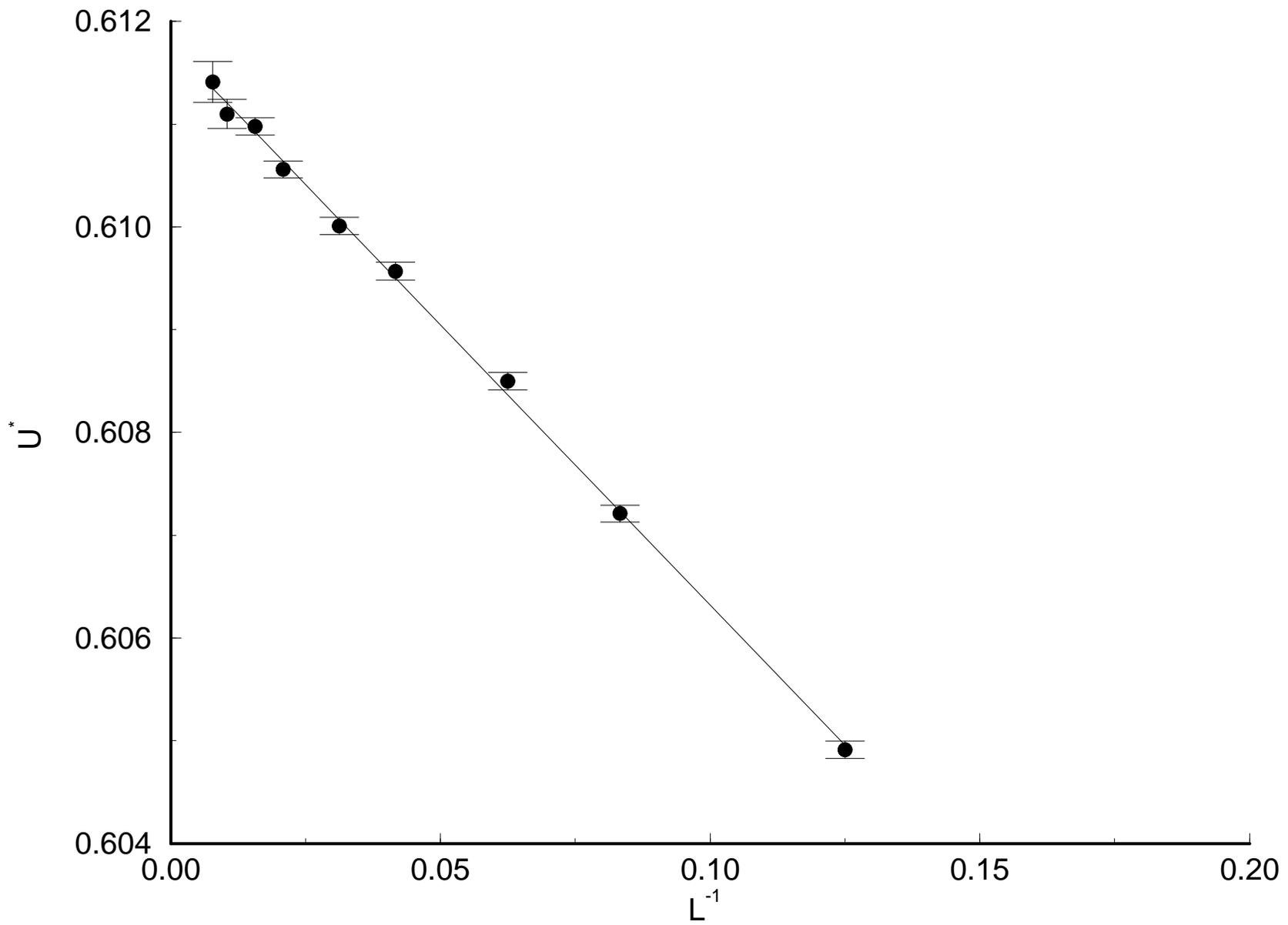
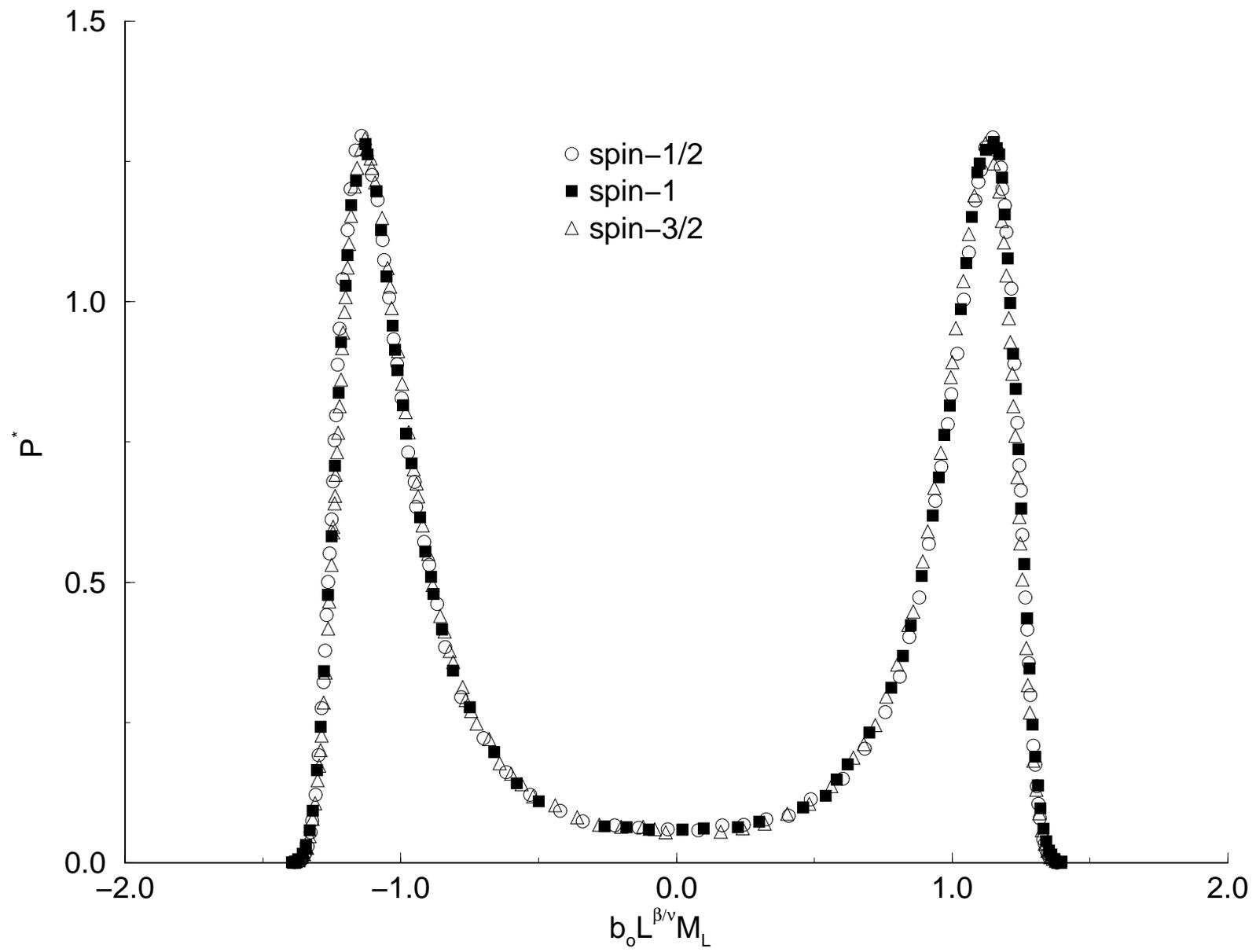
\end{document}